\documentclass[aps,twocolumn,prb]{revtex4}
\usepackage{graphicx}
\usepackage{amsmath}
\usepackage{amssymb}
\newcommand{\MCtwo}{Microtechnology and Nanoscience, MC2, 
Chalmers University of Technology, SE-412 96 G{\"o}teborg, Sweden}

\newcommand{\vdWDF}{{\mbox{\scriptsize vdW-DF}}}
\newcommand{\GGA}{{\mbox{\scriptsize GGA}}}

\newcommand{\CNT}{{\mbox{\scriptsize CNT}}}
\newcommand{\tot}{{\mbox{\scriptsize tot}}}
\newcommand{\nl}{{\mbox{\scriptsize nl}}}
\newcommand{\maxa}{{\mbox{\scriptsize max}}}
\newcommand{\DNA}{{\mbox{\scriptsize DNA}}}
\newcommand{\filled}{{\mbox{\scriptsize filled}}}
\newcommand{\hollow}{{\mbox{\scriptsize hollow}}}

\begin{document}

\title{A van der Waals density functional mapping of attraction in DNA dimers}

\author{Elisa Londero}\affiliation{\MCtwo}
\author{Per Hyldgaard}\affiliation{\MCtwo}
\author{Elsebeth Schr{\"o}der}\email{schroder@chalmers.se}\thanks{Corresponding author}\affiliation{\MCtwo}

\date{April 6, 2013}

\begin{abstract} 
The dispersion interaction between a pair of parallel DNA double-helix structures 
is investigated by means of the van der Waals density functional (vdW-DF) method.
Each double-helix structure consists of an infinite repetition of one B-DNA coil with 10 base pairs. 
This parameter-free density functional theory (DFT) study illustrates the initial step in a proposed
vdW-DF computational strategy for large biomolecular problems. The strategy is to first perform 
a survey of interaction geometries, based on the evaluation of the 
van der Waals (vdW) attraction, and then limit the evaluation of the 
remaining DFT parts (specifically the expensive study of the kinetic-energy repulsion) to the thus 
identified interesting geometries.  Possibilities for accelerating this second step is detailed in 
a separate study.  For the B-DNA dimer, the variation in van der Waals attraction is explored at relatively 
short distances (although beyond the region of density overlap) for a 360$^{\circ}$ rotation. This study 
highlights the role of the structural motifs, like the grooves, in enhancing or reducing the vdW 
interaction strength.  We find that to a first approximation, it is possible to compare the DNA double 
strand at large wall-to-wall separations to the cylindrical shape of a carbon nanotube (which is almost
isotropic under rotation).  We compare our first-principles results with the atom-based dispersive 
interaction predicted by DFT-D2 [J.~Comp.~Chem. \textbf{27}, 1787 (2006)] and find agreement in 
the asymptotic region. However, we also find that the differences in the enhancement that occur at shorter 
distances reveal characteristic features that result from the fact that the vdW-DF method is an 
electron-based (as opposed to atom-based) description. 
\end{abstract}

\maketitle

\section{Introduction} 

The key characteristics of a working biomolecular system is an
enormous richness of structural complexity and a robust working principle,
molecular recognition, for identifying geometries that optimize the
intermolecular binding and alignment. Stronger covalent or ionic binding 
determines the structure inside the molecules and provides
resilience towards statistical fluctuations.\cite{SchrodingerLife}
This permanence is, for example, of utmost importance in the preservation of the 
information contained in our deoxyribonucleic acid (DNA) 
genome.\cite{FrancisAndCrickSet,RutgersGenome} The binding that is of relevance for 
life processes, e.g., the molecular-recognition matching\cite{MolRegBook,MolRecognition1,MolRecognition2} 
of genes, is weaker to allow the reversible operations that Nature 
needs.\cite{SchrodingerLife} The molecular recognition arises as a delicate balance 
between steric hindrance, electrostatics, and van der Waals (vdW) attraction, the latter 
also termed the London dispersion interaction. 

The search for a deeper understanding of such biomolecular operation motivates development 
of a parameter-free computational theory that both provides transferability and  
computational efficiency.\cite{dna1,dna2} The structural complexity of the biomolecular systems
implies that the supramolecular system may express itself in a multitude of ways, and 
it is not certain that a given empirical (or semi-empirical) interaction 
model remains applicable for all emerging configurations and for varying charging states.

Density Functional Theory (DFT) is a highly valued condensed-matter theory 
tool that is our work horse in predictive computational theory of traditional 
materials problems.\cite{BurkePerspectives} In such systems,
the electron density remains high between the atoms, for example, in a bulk structure.
DFT also works excellently for individual molecules but it has until recently lacked an
account of the truly nonlocal correlations that underpin vdW interactions
between constituents in a molecular system. The issue is that molecular systems are inherently 
sparse:\cite{HardNumbers,ijqc,langrethjpcm2009} they must also contain low electron-density regions between
the molecular fragments. The sparseness is even the defining quality when Nature puts
molecular recognition to work among biomolecules.  Nevertheless, the last decade 
have seen developments that position DFT to overcome this previous limitation. 

The vdW density functional (vdW-DF) method\cite{Dion,Thonhauser,vdWDF2} 
has a track record\cite{langrethjpcm2009} as a general-purpose nonempirical DFT
that can characterize interactions in sparse and soft biomolecular systems.
The vdW-DF method accounts for vdW interactions by introducing true nonlocality in 
the density functional.  Being built as a physics-based description 
and within a constraint-based design, it has the electron-based 
description that helps transferability.
This is, for example, important for systems where it is essential to 
describe the vdW binding or attraction for several typical 
separations.\cite{MolCrys,MolCrys2,ourharris} This is true even if the plasmon-type
description may not capture the full complexity 
in the description of the far-apart-regime for some systems that effectively behave as
a low-dimensional metal.\cite{EarlyLength,DobsonReview,JPlength} 
There exist efficient implementations also for self-consistent 
evaluations. One such algorithm\cite{soler} uses a fast Fourier transform approach 
which accelerates evaluation for medium-to-large size systems while real-space 
evaluation approaches\cite{KGraphite,MolCrys2,Gulans,noloco,Junolo}  
may hold advantages for very-large-scaled systems, as discussed in 
Ref.\ \onlinecite{MolCrys2}.

This paper reports that a nonempirical vdW-DF characterization of 
the variation in the vdW attraction among biopolymers is indeed feasible. 
In vdW-DF, this vdW attraction is reflected in a nonlocal correlation term $E_c^\nl$. 
Other terms, including a single-particle kinetic energy term $T_s$ and the semilocal 
parts of the exchange-correlation energy $E_{xc}^0$ also contribute to the
vdW-DF total energy $E_{\tot}^{\vdWDF}$ but (for objects that are effectively neutral) 
it is the vdW attraction, contained in $E_c^\nl$, that dominates when the density 
overlap between fragments can be ignored.  Our vdW-DF characterization 
presented here thus computes
the $E_c^\nl$ variation to map the morphology dependence in the vdW attraction between
a pair of parallel, infinitely-repeated double-stranded (ds) DNA helices.
We find that the vdW interaction of such  DNA dimer 
is sensitive to the alignment of the major DNA structural 
motifs, particularly different orientations of the major and minor grooves of 
the two ds-DNA macromolecules.  We also identify and discuss a set of systematic 
changes that arise in the vdW attraction when the problem is investigated across
a range of interaction distances.\cite{MolCrys,kleis_prb2008} 
We further compare the $E_c^\nl$ variation to 
the evaluation of the semiempirical vdW term of a vdW-extended DFT method,\cite{DFTDset}
namely DFT-D2.\cite{Grimme} We use and extend analytical interaction 
results\cite{concentric,CNTparallelconcentric,kleis_cms2005}
that apply for the far (but not asymptotically far) regime, to illustrate the difference 
in nature between our electron-based vdW-DF study and the atom-based DFT-D approach.  
Moreover, we contrast this interpretation of the vdW attraction in a
DNA dimer with corresponding results for a dimer of carbon nanotubes (CNTs).

With this study we propose a vdW-DF computational strategy for large-biomolecules,
the strategy \textit{begins\/} non-empirical studies of the interactions by 
a mapping of the vdW-attraction term $E_c^\nl$. This is suggested to 
optimize computational efficiency. We observe that the $E_c^\nl$ mapping of
vdW attraction can today be carried out at the cost of about ten wall time 
minutes per DNA dimer 
interaction geometry. That is much smaller than the time it takes to evaluate 
the variation in the kinetic energy $T_s$ because of an excellent scaling of 
real-space evaluations of $E_c^\nl$, Ref.~\onlinecite{MolCrys2}. The cost
of computing $T_s$ through wavefunctions is the same for DFT-D and vdW-DF
so our suggested  strategy could also be relevant, for example, for large-system
DFT-D studies.\cite{DFTDset,Grimme} The motivation for our proposal is that a complete 
nonempirical vdW-DF evaluation can be focused on the relevant interaction morphologies 
that the $E_c^\nl$ mapping identifies. The possibilities for using an 
adapted Harris scheme\cite{harris,foulkes,nikulin,gordonkim} to accelerate 
the complete vdW-DF evaluation is discussed and assessed in a 
closely related publication, Ref.~\onlinecite{ourharris}. 

The  paper is organized as follows.  The ds-DNA and DNA dimer 
systems are described and discussed in Section II, which also introduces
the particular CNT structure investigated here.
In Section III we present the 
computational method used.  Section IV consists of a study of the 
nonlocal correlation energy variation as the distance between the 
DNAs is changed and the two structures are rotated.
It also contains comparisons of our vdW-DF-based results in the far 
(nearly asymptotic) region to the dispersion interaction predicted by 
DFT-D2,\cite{Grimme} and with results of similar studies for CNTs.

\section{Genome structure, a DNA model}

The primary structure of DNA is made up by four different nucleobases,
the adenine (A), guanine (G), cytosine (C) and thymine (T) bases, each 
covalently linked to a sugar.  
These building blocks are joined to a strand by phosphate groups connecting 
the sugars.  The two DNA strands form a 
double helix with (predominantly) hydrogen bonds linking base pairs,
thus forming the DNA secondary structure.
The repeated pairing between a purine base (A or G) and a smaller 
pyrimidine (T or C) produces a constant diameter for the double helix.  

\begin{figure}[t]
\begin{center}
\includegraphics[width=0.47\textwidth]{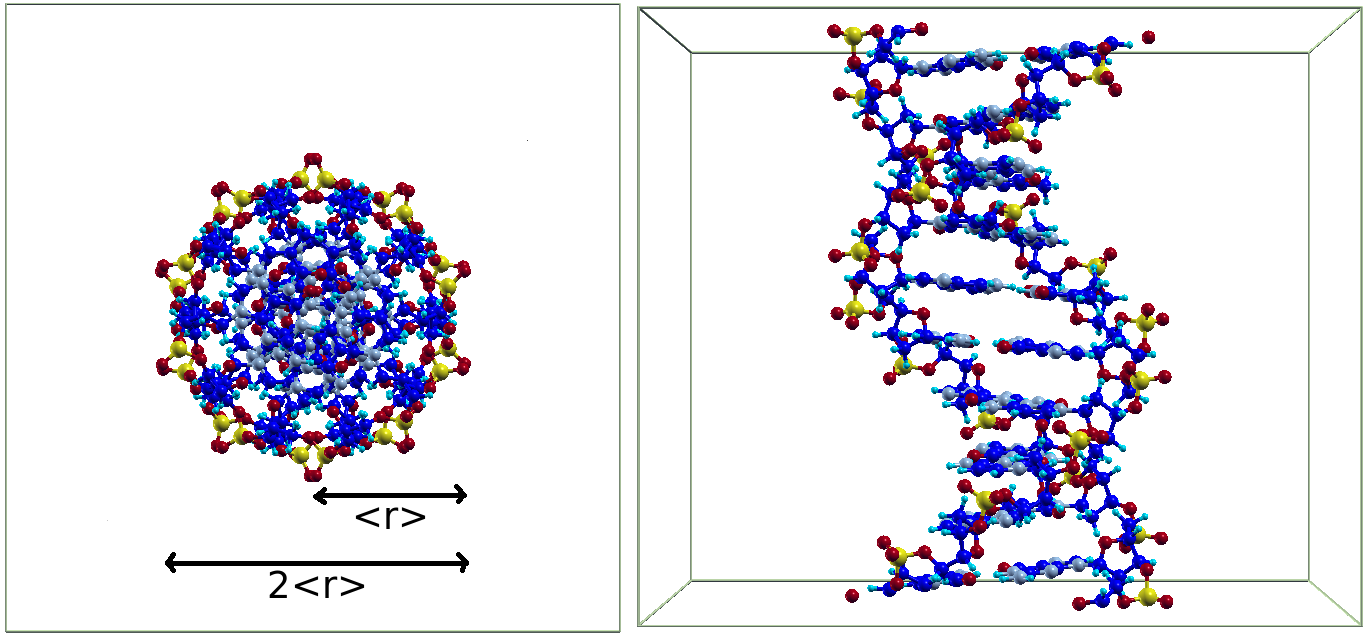}
\caption{\label{fig:X1}Along-axis and side views of a minimal 
repeat-unit cell model of the ds-DNA macromolecule. 
The atoms are P (yellow), O (dark red),
H (cyan), N (blue), and C (gray). 
Figure created using \textsc{XCrySDen}.\cite{xcrysden}
}
\end{center}
\end{figure}

One of us was previously involved in a vdW-DF study of the nucleobases,
focusing on the stacking interactions.\cite{dna1} Even 
if the vdW forces by themselves may have limited selectivity 
(for small molecules like base pairs), the vdW forces are essential in the 
building of this secondary structure because they drive, optimize, 
and hold a structural assembly until more stable bonds can be formed.  
This mechanism is fundamental in the case of molecular recognition and 
in DNA replication.  The previous study\cite{dna1} provides vdW-DF 
analysis and results for the Rise and Twist of the base pairs inside 
(that is, between the backbones of) the ds-DNA structure.  
The nonlocal correlation contribution $E_c^\nl$ to vdW-DF will by itself 
have difficulties in aligning small molecules but the vdW forces still 
position these components (base pairs) close enough that the corrugation 
from kinetic-energy repulsion is expressed. The previous vdW-DF 
study\cite{dna1} reported a very good agreement between vdW-DF 
characterizations and experimental information (for example, correlations in 
base-pair rotation angles as extracted from the Nucleic Acid 
Database\cite{RutgersGenome}).  This progress highlights possibilities 
of a broader application of the vdW-DF method to soft matter 
challenges and in particular to refine the description of the 
vdW attraction between biopolymers.

We focus our large-scale vdW-DF interaction study on dimers of a 
ds-DNA model structure in the B-form.
Both ds-DNA fragments are assumed to have the periodically repeated 
sequence of nucleobases GCAATACGGT.  The ds-DNA atomic structure is built from 
idealized coordinates.\cite{SB1,SB2,Jason} In our model the axes of the 
two ds-B-DNA fragments are straight and parallel to each other.

Figure \ref{fig:X1} shows the atomic positions in the minimal-repeat 
unit cell for one ds-B-DNA molecule.
On the most coarse level it reflects a cylinder-like form for DNA
but it also 
shows both major and minor grooves of the double helix.  This DNA model contains 635 atoms 
and is limited to one coil, but it is periodic and 
is used as the repeat unit cell of a DFT calculation so 
that the DNA is still described as a macromolecular system (through the 
infinite repetition of the unit cell along the DNA axis).  

The polar sugar-phosphate strand 
on the external part of the chain can form favorable interactions with ions 
in a solvent.  Each phosphate group then carries a negative charge which is 
balanced by positive ions, the counter ions, in the surrounding liquid. 
The counter ions stabilize the structure, making the DNA-ion system charge 
neutral.  Correlation between the concentrations in the counter ions may 
themselves contribute further to an attraction between the DNA-ion system 
and another organic molecule. 
However, we shall not consider such effects here.
Instead we investigate the vdW binding arising from nonlocal 
electron correlation effects between two ds-DNA structures 
assuming (and enforcing in our DFT calculations) that the DNA 
repeated-unit cell model structure itself remains charge neutral.  
A forthcoming study addresses the effects on the vdW attraction
of the charging by counter ions.

The double-helix structure is very regular, and when viewed
along the axis appears almost isotropic
(left panel of Fig.~\ref{fig:X1}). 
This average isotropy of DNA simplifies our study 
of the sensitivity of the vdW binding to macromolecular structural motifs.
The average isotropy also helps in the comparison with CNT dimer 
interactions as we can define an approximate 
``radius" of DNA. We can thus compare the vdW interactions as a function of 
wall-to-wall distances $\Delta$ in dimers of DNA and dimers of CNT, using
the analysis of Refs.~\onlinecite{concentric,CNTparallelconcentric,kleis_cms2005,kleis_prb2008}.
For DNA we use the radius that is defined as the average of  the 
helix backbone P and O atoms distances 
from the DNA center (8.5 {\AA} and 9.7 {\AA}). This yields 
the approximate DNA radius $\langle r \rangle_\DNA \approx$ 9.1 {\AA} 
(shown by a double-arrow line in Figure \ref{fig:X1}), 
close to that of the standard
literature description where the diameter is taken to be $\approx 20$ {\AA}.
For a given                                                              
center-to-center separation $d$ of the DNA dimer structure we thus use
$\Delta=d-2\langle r\rangle_\DNA$ as the wall-to-wall separation 
of the DNA dimer.

The approximate similarity of DNA with the cylinder form of a  
CNT represents one major structural motif of the structure. 
Another major structural motif is obviously the existence of 
grooves (right panel), a feature which is distinctly different from 
what characterizes the CNTs.

The systems studied here are sparse\cite{langrethjpcm2009}
so that the vdW interaction (with electrostatics) dominates in the
intermolecular regions of low electron density. 
By basing the function 
on weaker forces, Nature ensures a truly remarkable resilience of our genetic
code.\cite{SchrodingerLife,MolRecognition2}

\begin{figure}[t]
\begin{center}
\includegraphics[width=0.47\textwidth]{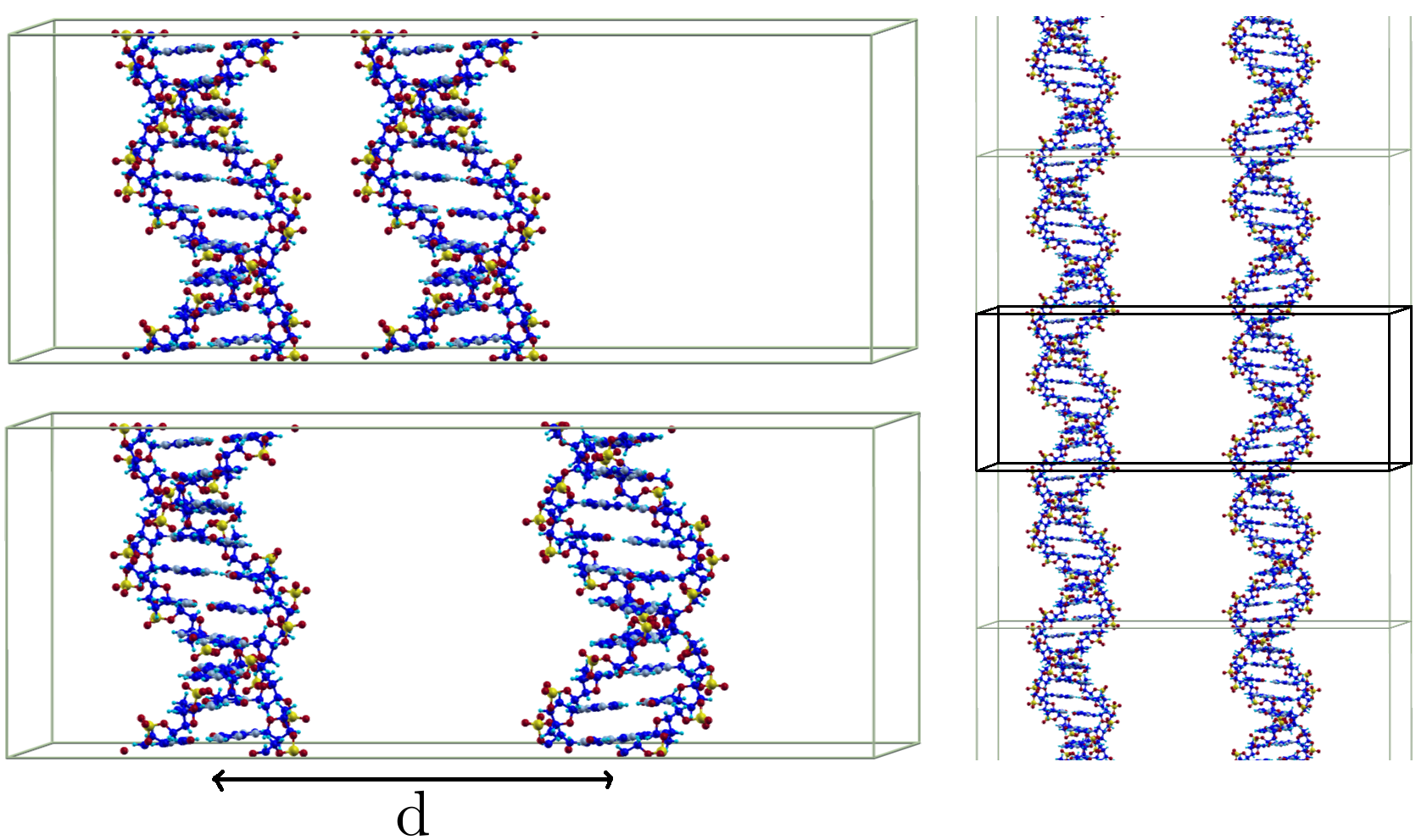}
\caption{\label{fig:X2} Left: Example of a system obtained starting 
from a single 33.8 {\AA} period of the ds-DNA that is copied, 
translated and rotated inside an enlarged unit cell.
Right: Portion of the final system studied. Here we have piled a 
number of unit cells (not all shown) on top of each other in 
preparation of the real-space evaluation of 
Eq.~(\protect\ref{ECNLexpress}).
Figure created using \textsc{XCrySDen}.\cite{xcrysden}
}
\end{center}
\end{figure}

\section{Computational details}

DFT calculations of interacting systems are effectively limited by the 
computational challenge of accurately calculating the kinetic energy, a 
step which requires solving the noninteracting particle Schr{\"o}dinger 
equation in three dimensions.  One could fear that the nonlocal nature 
of the correlations that define the dispersive forces, i.e., the vdW-DF 
evaluation of the $E_c^\nl$, may also represent a computational challenge 
for large systems. However, efficient algorithms are now in place and there
are, effectively, no other computational bottle\-necks for biomolecular
vdW-DF and/or DFT-D studies, like those presented here, than the 
evaluation of the (noninteracting) kinetic energy $T_s$.

We note that since $E_c^\nl$ is far less costly (for large systems) 
than any calculation of eigenstates, it is wise to \textit{begin\/} a 
first-principle biomolecular interaction study by mapping out the 
$E_c^\nl$ variation; this saves the costly determination of the kinetic-energy 
repulsion to relevant interaction morphologies. Here we pursue this first mapping 
step of evaluating $E_c^\nl$ to identify optimal DNA-dimer 
configurations. We also characterize the sensitivity of the vdW binding 
to variations in alignment of DNA structural motifs. 

Figure \ref{fig:X2} illustrates our process in our $E_c^\nl$-mapping. 
It is formally the first step in an overall Harris-type approach for an 
accelerated vdW-DF evaluation of inter-molecular
interactions.\cite{ourharris}

The DNA-dimer electron density $n_d$ is needed for evaluating $E_c^\nl[n_d]$ for the dimer.
For the individual DNA we calculate the electron density in a Linear Combination
of Atomic Orbitals (LCAO) characterization, available in the DFT code GPAW.\cite{gpaw}
As an approximation of  
$n_d$ we use the superposition 
\begin{equation}
n_d(\mathbf{r})=n^\GGA_1(\mathbf{r})+n_2^\GGA(\mathbf{r})
\label{eq:nd}
\end{equation}
of two copies of the individual density, $n^\GGA_{1,2}$.  
The superscript indicates a standard self-consistent GGA calculation.

The calculation of the electron density that makes up $n^\GGA_{1,2}$ is carried out  
at the $\Gamma$ point.
The length of a period along the ds-DNA is 33.8 {\AA}, and for 
determining the electron density we use an orthorhombic unit cell of size 
$39.5\times 40.4\times33.8$ {\AA}$^3$. The electron density is described on 
a grid with a spacing of approximately 0.20~{\AA}.

The extended system for the superposition $n_d$ is created by first 
enlarging the unit cell of the individual DNA by factors 
$2.3\times 2.3\times 1$ to obtain a unit cell of size 
$90.9\times 92.9\times 33.8$ {\AA}$^3$ (in Figure \ref{fig:X2} the original
unit cell is enlarged only along the first axis, for visual reasons).
Then the density for the second ds-DNA is added, appropriately translated 
and/or rotated around the DNA axis. 
For rotated ds-DNA the values of the electron density on the spatial grid are 
interpolated from the grid values before rotation.  

The set of leftmost panels of Fig.~\ref{fig:X2} summarizes this  
step in our vdW-DF study.  By also rotating the ds-DNA densities we can map 
out various mutual alignments of the groove structure and study the ramifications for
the vdW attractions (as described in vdW-DF or in DFT-D).

For each of the considered separations $d$ of the dimer and for each of 
the relative angles of the fragments we evaluate $E_c^\nl$ by carrying 
out the integral  
\begin{equation}
E_c^\nl[n]=\frac{1}{2}\int 
n(\mathbf{r}_1)\phi[n](\mathbf{r}_1,\mathbf{r}_2)n(\mathbf{r}_2)
d\mathbf{r}_1 d\mathbf{r}_2 
\label{ECNLexpress}
\end{equation}
for the superposition density $n_d$ of Eq.\ (\ref{eq:nd}). Here 
$\phi[n]$ is given from a 
tabulated kernel function, but $\phi[n]$ still reflects the overall
density variation in the interacting components.\cite{Dion,ourharris,newvdWanalysis}
Our code for the postprocess calculation of the $E_c^\nl$ value is a 
real-space code.
For the periodicity along the DNA axes
we perform the real-space
evaluation step by replicating the DNA density along the $z$ direction 
(the DNA axis), for a total length of up to 540 {\AA} (depending on the 
need in each calculation) and then evaluate (\ref{ECNLexpress}) with one
spatial coordinate restricted to the central unit cell, as described
in Ref.~\onlinecite{MolCrys2} and illustrated also in 
Ref.\ \onlinecite{adenine}. This is  indicated by the                                                     
right panel of Fig.~\ref{fig:X2}

The ds-DNA structure can be roughly
approximated by a CNT-like cylindrical structure
(although filled) when viewed at large wall-to-wall
separation.
For a comparison we present 
CNT calculations using a (15,15) CNT with the electron density (of
the individual CNT) obtained in a
$40.7\times40.7\times2.46$ {\AA}$^3$ unit cell that contains 60 atoms.
The radius of the (15,15) CNT is $\langle r\rangle_\CNT$=10.0~{\AA},
approximately the size of the DNA radius. 
For the CNT we have chosen to use the GPAW finite differences mode 
(instead of LCAO) as a basis for the individual CNT electron density 
variation. We also set a grid for the description of the electron 
density that has a spacing less than 0.135 {\AA} between grid points, 
and we use a Monkhorst-Pack $k$-point sampling of the Brillouin zone with a  
1$\times$1$\times$16 mesh. Like for the DNA dimer, the CNT dimer density 
is obtained as a superposition of the individual densities.

In an additional set of comparisons, we also report DFT-D2 calculations 
of the vdW attraction for dimers of DNA and
for dimers of CNTs. These are based on the dimer atomic configurations. 
The DFT-D calculations are carried out as follows. For each of the DNA/CNT atoms 
in the original unit cell, we sum the pair contributions over atoms in the
repeated copies of the DNA or CNT unit cell, with the repetitions extending
over 473 {\AA} (or 8890 atoms) for the DNA-dimer problem and over 404 {\AA} 
(or 9900 atoms) for the CNT-dimer problem.

\section{Results and discussion}

\subsection{A mapping of mutual vdW attraction in a soft twinning of 
two B-DNA coils}

\begin{figure}[t]
\begin{center}
\includegraphics[width=0.45\textwidth]{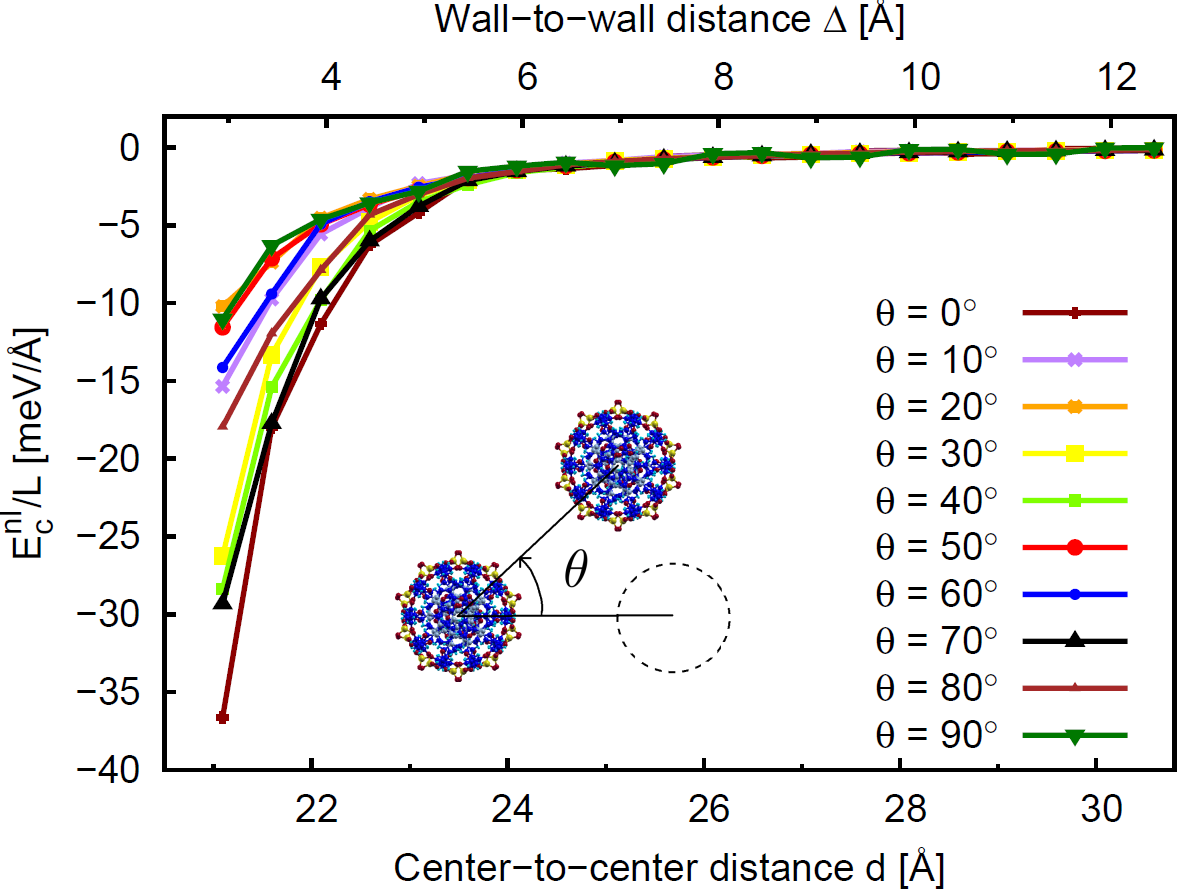}
\caption{\label{fig:X3}Nonlocal correlation energy per length $E_c^\nl/L$ 
of the DNA dimer as a function of rotation angle $\theta$ and separation.
Both the center-to-center $d$ and wall-to-wall distances $\Delta$ are 
shown in the plot on the bottom and top horizontal axis, respectively.
The picture in the insert shows the geometry of the system as the angle 
$\theta$ is varied.
}
\end{center}
\end{figure}

We first explore the vdW sensitivity of the DNA interactions  
as one DNA is rotated (rigidly) around another DNA, as illustrated in the insert of 
Fig.\ \ref{fig:X3}.  In the rotation $\theta$ the individual DNA is not rotated 
around its own axis. Instead the situation corresponds to keeping the DNA dimer axes 
fixed in space and rotating the left hand side DNA by the angle $-\theta$ around its own 
axis while simultaneously rotating the right hand side DNA around its 
own axis by the angle $-\theta$.  Effectively we are thus exploring what
variation arises in the vdW attraction effects when twinning the two B-DNA coils.
We restrict this twinning to situations that create no significant density overlap.

Each of the curves in Fig.\ \ref{fig:X3} corresponds to a certain 
relative twinning rotation $\theta$ and shows the 
interaction $E_c^\nl$ at various dimer separations (wall-to-wall distance 
$\Delta$ or center-to-center distance $d$) from values smaller than 
expected at the dimer binding, $\Delta= 2.94$ {\AA}, and up to 
$\Delta=12.4$ {\AA}.  

We note that the value of $E_c^\nl$ varies strongly with wall-to-wall separation $\Delta$,
but except at the very small separation $\Delta= 2.94$ {\AA},
the order of the curves of largest and smallest contribution remain the same.

We cover the variation in the rotation angle $\theta$ up to 90$^{\circ}$ 
with steps of 10$^{\circ}$, and we find variation in the DNA vdW binding 
$\theta$.  At about $\Delta=3.7$ {\AA} (within binding-distance range) 
the variation, per unit length, in 
$E_c^\nl$ with $\theta$ amounts to about 7 meV/{\AA}.

There are several local minima 
in the variation when viewed at fixed distance $> 3$ {\AA} 
(at $\theta= 0^{\circ}$, $40^{\circ}$, and $70^{\circ}$).  
The variation in the attraction with $\theta$ is correlated with the 
alignment of the grooves, but the presence of several local minima also 
shows that detailed structure in the DNA further influences the interaction. 
For example, the DNA is not a continuous double helix, but consists of a 
series of base pairs and the associated parts of the back bone, at
3.4 {\AA} apart along the DNA axis, and the choice of bases in the base 
pairs varies along the DNA. Thus we should not expect smooth, largely monotonous
curves when the relative orientations of the DNA fragments are changed.
 
\subsection{Interaction effects by the alignment of motifs}

\begin{figure}[t]
\begin{center}
\includegraphics[width=0.45\textwidth]{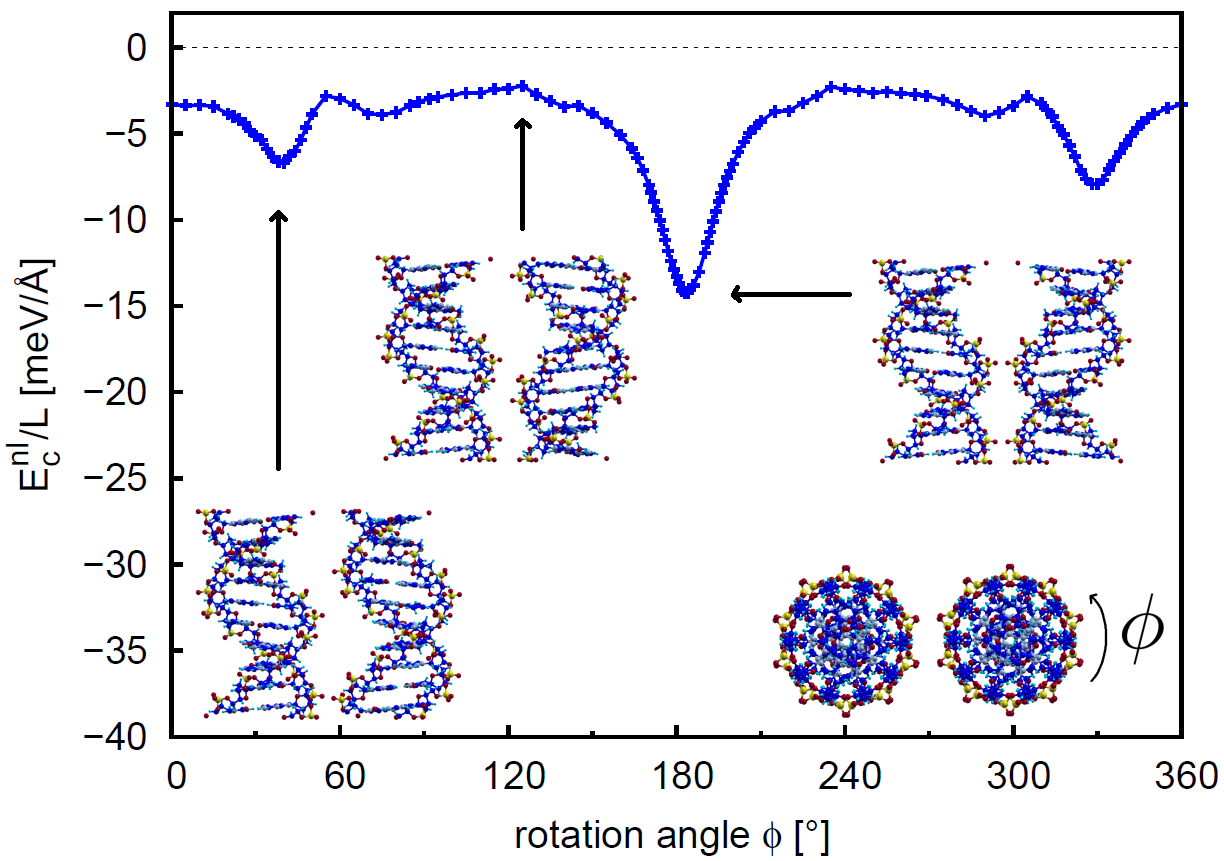}
\caption{\label{fig:X6}
The vdW binding as a function of the rotation $\phi$ of one of the DNA
molecules around its axis while keeping the other at a fixed orientation.
The separation of the dimer is $\Delta=4.5$ {\AA}.}
\end{center}
\end{figure}

The curves in Figure \ref{fig:X3} show significant variation 
in $E_c^\nl$ with relative orientation of the DNA fragments.
We therefore further explore the sensitivity to alignment
of structural motifs by keeping one 
of the two DNA fragments fixed and rotating the other  
by an angle $\phi$.
Our study creates a 360$^{\circ}$ mapping of $E_c^\nl/L$ 
versus $\phi$, the rotation of one B-DNA coil. The results
are shown in Figure \ref{fig:X6}. 
The interaction is evaluated at a separation $\Delta= 4.5$ {\AA}
which is slightly larger than the expected binding distance.
At closer distances some orientations will result
in overlap of high densities at the outer O atoms on
the two fragments (because DNA is not truly cylindrical).

Fig.\ \ref{fig:X6} demonstrates 
a significant and rapid variation of the vdW binding with the alignment 
of motifs, i.e., the alignment or disalignment of the major and minor groves.   
There is an approximate symmetry of the nonlocal correlation energy around 
the central drop (at 183$^{\circ}$), with the value $-14.2$ meV/{\AA}, 
or 12.0 meV/{\AA}
lower in energy than the least attractive orientations at that separation,
namely at 125$^{\circ}$ and 235$^{\circ}$.

The $E_c^{\nl}$ (or vdW attraction) minimum at 183$^{\circ}$ corresponds to 
the alignment of the two structures in a way that, for each of the fragments considered, two of 
the outermost O atoms face each other directly.
This also means that the minor and major groove of the first structure are
aligned with the corresponding grooves on the second DNA, as
depicted in the inset picture associated to this point.

The two orientations (at $\phi=125^{\circ}$ and 235$^{\circ}$) with the largest
$E_c^{\nl}/L$ values have a vdW attraction of only $-2.4$ meV/{\AA}, respectively 
$-2.6$ meV/{\AA}, compared to the situation of the two DNA fragments being far apart.
This orientation  has a large amount of vacuum between the two
structures along their entire length.

Two other (minimum) features correspond to a clear enhancement in 
attraction, even if more modest. 
These local-minimum angles appear as approximately symmetric with respect 
to the global minimum in the $\phi$ variation. They appear at 40$^{\circ}$ 
and 328$^{\circ}$ with vdW attraction
$-6.7$ meV/{\AA} 
and with
$-7.9$ meV/{\AA}. 
This is
4--6 meV/{\AA} lower in energy compared to the least attractive orientation.
These two local minima correspond to an 
alignment of the two DNA in which only one of the external O atoms 
faces another O atom in the other DNA.

Altogether, our vdW-DF evaluation thus makes it possible to detect 
variations in the nonlocal correlation energy within a $\sim$12 meV/{\AA} 
broad range due to the alignment of the structural motifs.

\subsection{A comparison with vdW-extended DFT and with a CNT dimer}

\begin{figure}
\begin{center}
\includegraphics[width=0.45\textwidth]{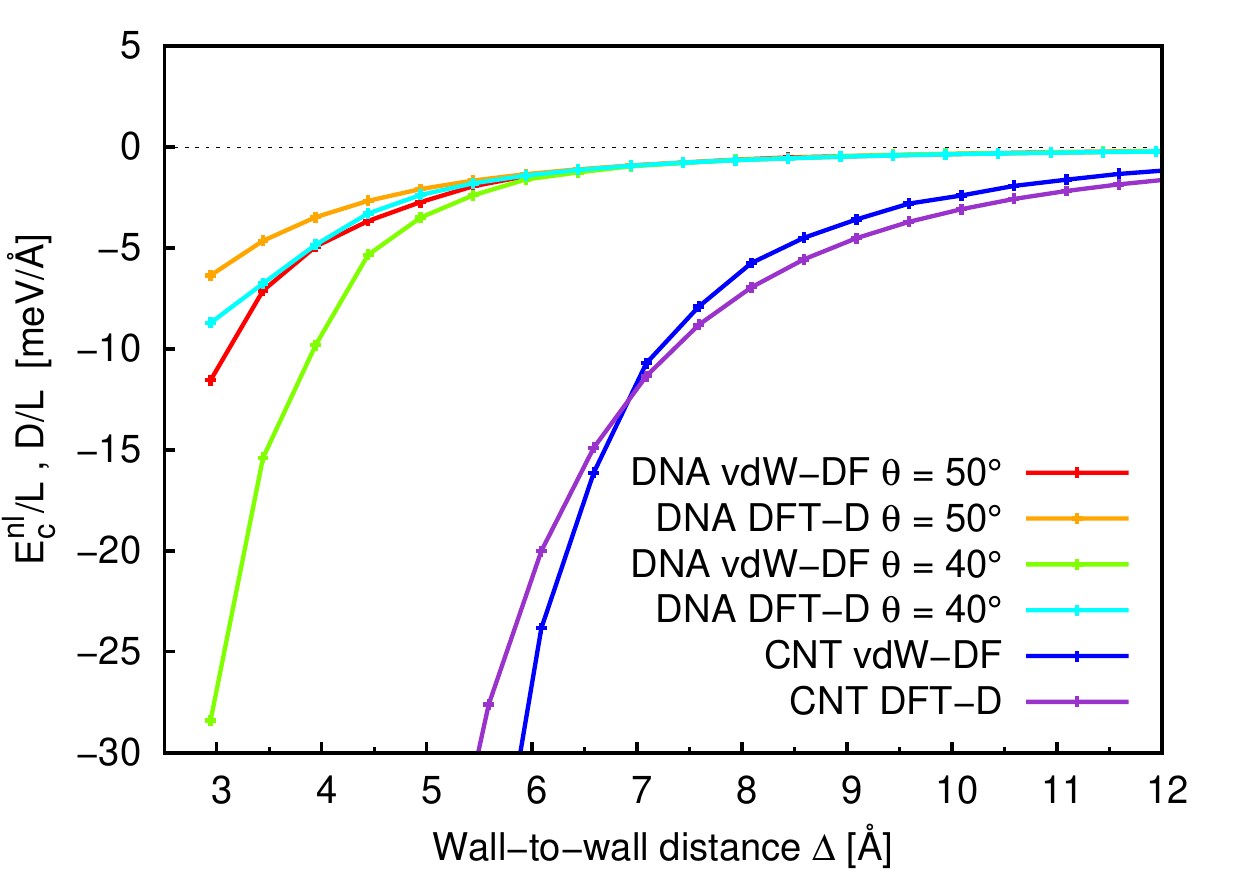}
\caption{\label{fig:X4}
DNA and CNT vdW interaction per length as a function of wall-to-wall 
distance $\Delta$.
}
\end{center}
\end{figure}

We continue the analysis by comparing and contrasting our vdW-DF results 
with results from a summation of pair contributions, using a traditional 
atom-based dispersive interaction form, here as provided by Grimme for the 
dispersion term $D$ of DFT-D in the version DFT-D2.\cite{Grimme}
Figure \ref{fig:X4} repeats the $\theta=40^{\circ}$ and 
$\theta=50^{\circ}$ curves for the DNA vdW interaction and includes also 
the curves for the dispersion term $D$ of DFT-D for the same two rotations.

For separations larger than about $\Delta=6$ {\AA}, we find reasonable 
agreement between the results of the vdW-DF and DFT-D procedures 
(Fig.\ \ref{fig:X4}). At 
smaller separations we expect the results to differ because at close 
range it becomes important that the dispersion interaction arises mainly 
in the valence electron region, not at the atomic centers, as assumed in
DFT-D. Indeed, we do see a difference in the results at small separations, 
with diminished attraction in the results from DFT-D compared to those
from the vdW-DF method.

We also compare in Figure \ref{fig:X4}  to the attraction
of a pair of parallel (15,15) CNTs that have approximately the same radius
as DNA, $\langle r \rangle_\CNT = 10$ {\AA}. The CNT data are
calculated with vdW-DF and DFT-D. The attraction of the CNTs is stronger than for
the DNA dimer. We note that there are about 30\% more atoms per length (relevant for
DFT-D, although also the species of the atoms matter) and more integrated 
electron density per length (relevant for vdW-DF) in the (15,15) CNT than 
in the DNA. In a previous study\cite{kleis_cms2005,kleis_jcp2005} of 
dimers of polyethylene (PE), polypropylene (PP), and polyvinyl chloride (PVC) 
two of us found a strong dependence of the vdW interaction (in a simplified
method) on the integrated amount of electron charge (per length) in the polymer. 
In a full vdW-DF calculation, including $E_c^\nl$, of PE in dimers and in a crystal,
we found\cite{kleis_prb2007} the vdW term $E_c^\nl$ to depend strongly on the
separation of the molecules and less, but not insignificantly, on the relative orientation. 

\begin{figure}
\begin{center}
\includegraphics[width=0.45\textwidth]{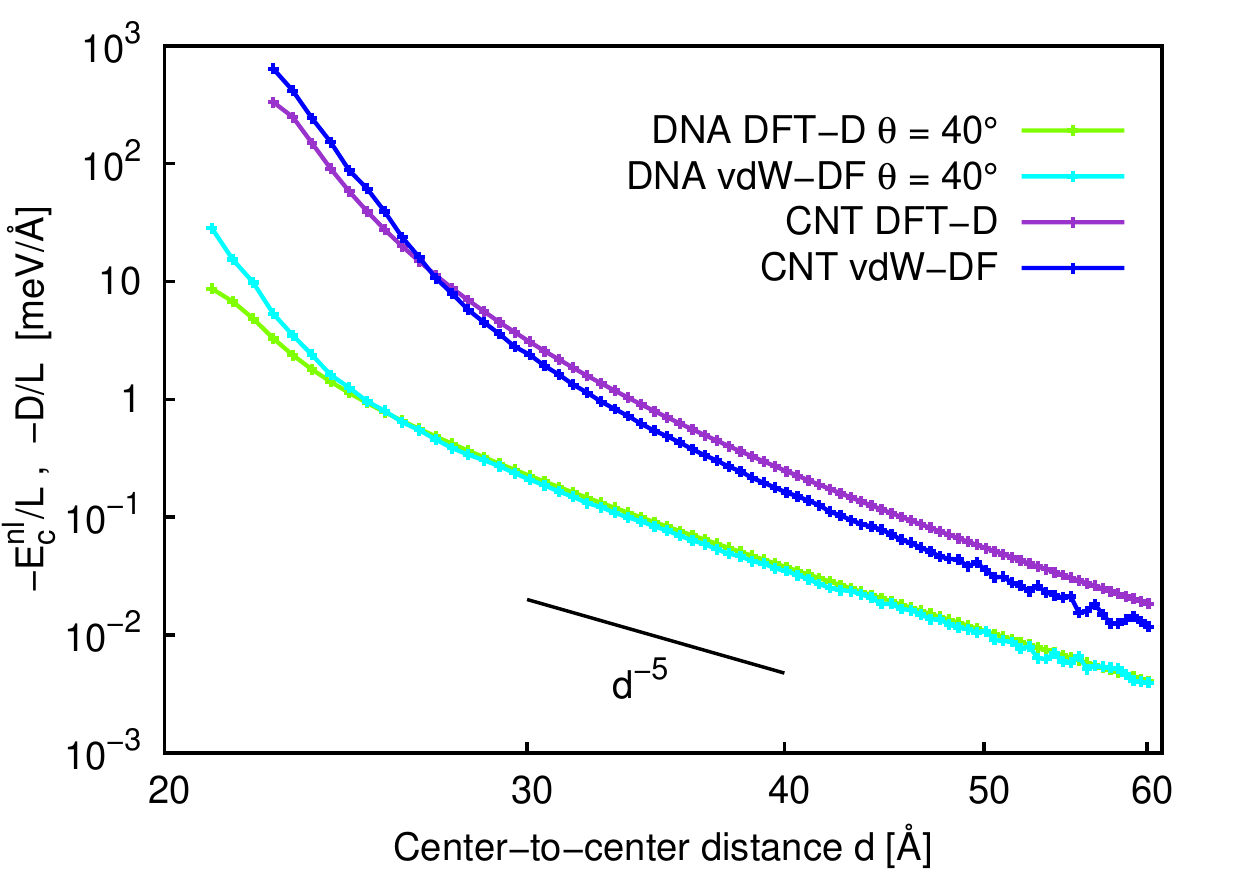} \\
\includegraphics[width=0.45\textwidth]{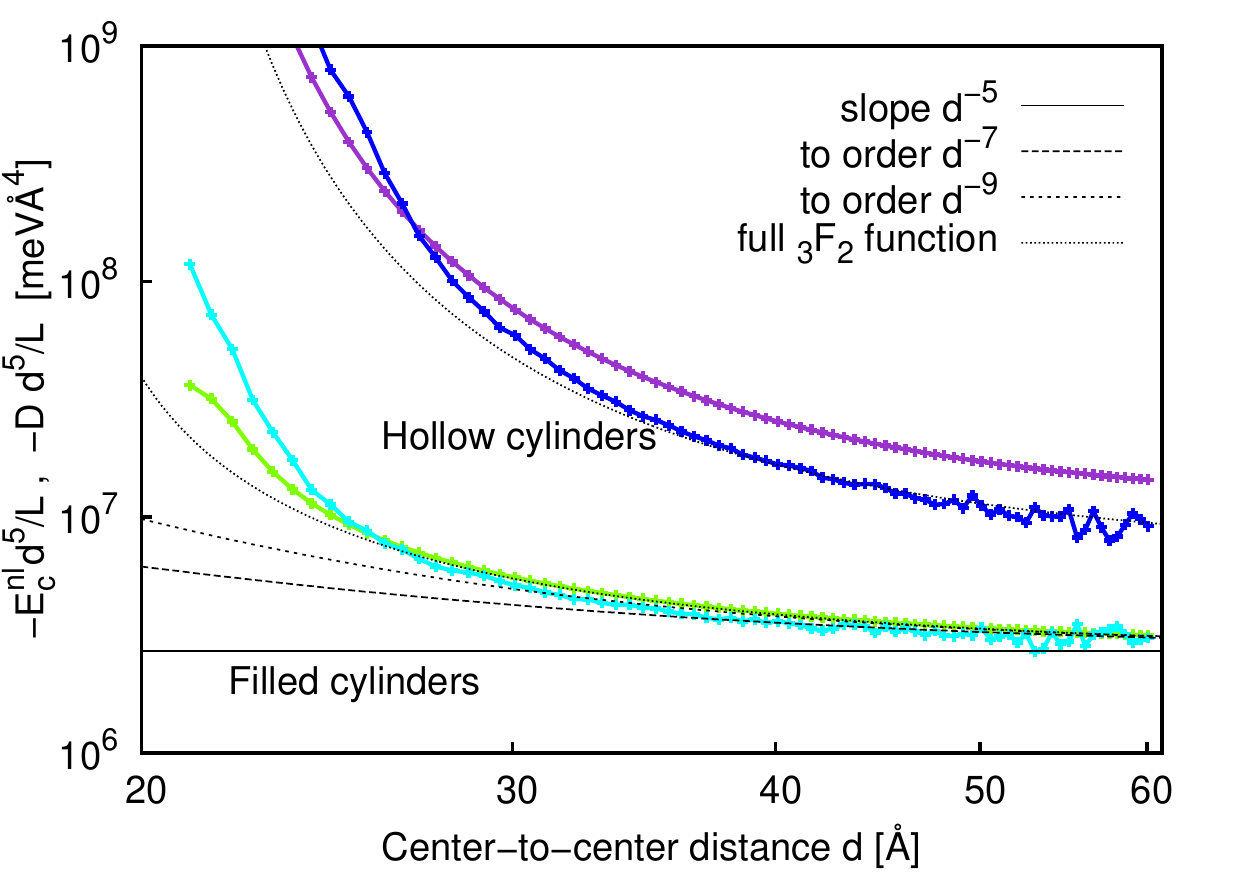} 
\caption{\label{fig:X5}
The vdW interaction in DNA and CNT dimers using vdW-DF and DFT-D.
}
\end{center}
\end{figure}

We stress that there are significantly larger differences between 
the estimates of the vdW attraction for a CNT dimer based on our 
vdW-DF evaluation and the DFT-D2 evaluation, 
than for a DNA dimer in Fig.~\ref{fig:X5}. 
Some of the difference must be ascribed to the difference that exists
in the underlying GPAW calculation of the single-fragment electron density (a courser-grained LCAO 
procedure for the DNA and a finite-different approach for CNT). 
However, we expect (and are presently testing in a continuation
vdW-DF study) that the differences mostly reflect the fact that for vdW-DF it is
important not only how many electrons are in a molecule but also how they are distributed in
space.\cite{newvdWanalysis} It is also the case that for vdW-extended DFT methods,
(which use parameters to specify the per-atom contribution to the vdW interaction strength) there is
now an awareness that sp2-hybridized carbon (the form that roughly applies for the CNTs) will need a different 
parametrization than carbon  in other organic molecules (like DNA).

\subsection{On the validity of a near-asymptotic interaction form}

Fig.\ \ref{fig:X5} analyzes the asymptotic forms of the vdW-DF and 
DFT-D2 results for the (15,15)-CNT and DNA interaction curves for 
the $\theta=40^{\circ}$ orientation.
The largest center-to-center distance $d=\Delta+2\langle r\rangle$ 
is here approximately 60 {\AA}, i.e.,
at the furthest separation $d_\maxa$ the ratio to the DNA radius
is $d_\maxa/\langle r \rangle_\DNA\approx 6$.
This means that only the points furthest out can possibly be considered 
truly asymptotic. 
In the asymptotic and almost-asymptotic range the $\theta=40^{\circ}$ 
and $50^{\circ}$ curves are identical and we therefore only show the
$\theta=40^{\circ}$ curves in this plot.
 
Beyond 
$d\approx 50$ {\AA} and up to $d_\maxa$, which is a small range here,
the interactions for the DNA and the CNT, both from vdW-DF and DFT-D calculations,
scale approximately as 
$-d^{-p}$ where $p=5$. This is the asymptotic scaling expected for parallel
cylinders. 
The slope of $-d^{-5}$ is drawn in the top panel of Figure \ref{fig:X5}
as a guiding line.
In general the interaction of parallel cylinders at equal radius $b$ 
at large (but not necessarily asymptotic) separation $d$ is given by the 
generalized hypergeometric function ${}_3F_2$, as described in the Appendix.
This applies for cylinders of a continuous material, which our systems with
atoms and varying electron distribution are not quite.

In the bottom panel of Figure \ref{fig:X5} we plot the interaction 
per length, times $d^5$. In the truly asymptotic region for
a pair of cylinders we expect a vanishing slope,
and for separations a bit smaller we expect the next terms in the expansions
(\ref{eq:expanfilled}) for filled cylinders and (\ref{eq:expanhollow}) 
for hollow cylinders to contribute. As seen
in the bottom panel of Figure \ref{fig:X5}, neither the DNA nor the CNT dimers 
are in the fully asymptotic region at $d<d_\maxa$ (the curves are not totally
flat). However,  after adding the first few terms of the expansions
(\ref{eq:expanfilled}) or (\ref{eq:expanhollow}) we find good agreement with 
the form of the generalized hypergeometric function~${}_3F_2$ outside the 
near-binding region, $\Delta = d-2\langle r \rangle  \lesssim 4$--10 {\AA}. 

Whether a pair of cylinders are filled or hollow does not
affect the lowest order term in the asymptotic expansion,
the $d^{-5}$-term. However, it does change by a factor of 2 the coefficient on the
next order term, the $d^{-7}$-term, and thus the interaction
curve in the not-quite-asymptotic region. When the DNAs or CNTs are close to
each other the
separation is too small for the expressions in terms of the
generalized hypergeometrical functions to be valid.

It is clear that DNA is not a regular cylinder structure and that the 
existence of grooves must play an essential role, as also seen for the 
interaction as a function of relative orientation of the DNA dimer. However,
far from the binding region, basing the description on an assumption of 
the electrons being distributed in a filled cylinder is a good approximation.

The analytical evaluation reflects the morphology of the interaction fragments
and is therefore a good approximation to the pair-potential summation that underpins
a DFT-D evaluation. Fig.~\ref{fig:X5} shows that the analytical description applies well as an 
approximation to the DFT-D results when the fragments are sufficiently removed 
to also make the above-stated assumptions meaningful. However, Fig.~\ref{fig:X5}  also shows that  
our $E_c^{\nl}$-based calculation of the vdW attraction maintains differences for the near-asymptotic
form out to further distances than does the DFT-D2-based evaluation. 

We ascribe these differences to the vdW-DF ability to include 
multipole and some collectivity effects through its plasmon-pole description\cite{Dion} and 
its emphasis on reflecting the electron-density variation.\cite{MolCrys,MolCrys2}

\subsection{Towards a full vdW-DF interaction study}

We note that at distances $\Delta \lesssim 4$ {\AA} it is also
important to include the remaining components of the vdW-DF method, see
Ref.\ \onlinecite{MolCrys}.

In a closely related study\cite{ourharris} 
we assess the possibility of using an adaption of the 
Harris scheme\cite{harris,foulkes,nikulin,gordonkim}
to accelerate the evaluation also of the remaining vdW-DF (or DFT-D) components,
predominantly seeking to bypass repeated evaluations of the kinetic energy variation
$T_s$. This study shows that a high degree of accuracy can be achieved. A forthcoming
paper will report the results for the DNA dimer problem and will detail 
effects that the DNA charging state might have on the vdW attraction.

\section{Summary and outlook}
In this paper, we show that the vdW-DF functional has the potential to describe 
extended systems with the accuracy of DFT, thus opening a way to the 
description of complex soft phenomena.

We find that our computational strategy for first-principle
vdW-DF characterization (with our moderate-level access to HPC) can give 
some qualitative and even semi-quantitative results for biomolecular 
systems even before the full vdW-DF (or for example DFT-D) evaluation of 
the kinetic-energy repulsion effects is completed.

An accelerated vdW-DF computational framework\cite{ourharris} study 
can be expected to work well for investigations of vdW bound systems.  
Here, we have illustrated the possibility of an even faster initial 
mapping by evaluating the vdW-binding component.
An additional component in the strategy is to adapt the ideas of the 
Harris scheme as explored in a 
parallel study.\cite{ourharris}
The motivation such strategy is easily stated:
even if our DNA model structure is only a model system of the full secondary
genome structure, the individual ds-DNA still contains 635 atoms and 
the dimer system contains 1270 atoms. The size of the problem prevents us 
(at the supercomputing facilities to which we have access) from performing an 
ordinary potential-energy calculation by means of a conventional DFT implementation.
More realistic studies will contain even more atoms per unit cell, or have other 
similar complications. Thus any computational simplification able to 
keep most of the original accuracy is welcome.
That observation is true whether or not we wish to pursue a vdW-DF or a DFT-D study.

The nonlocal correlation energy $E_c^\nl$, which contains the
essential information about the vdW binding, is accessible with a limited
computational effort. We have found that this $E_c^\nl$ evaluation has essentially perfect
scaling up to at least 2000 cores in our real space implementation.\cite{MolCrys2}

We are thus proposing to \textit{initialize} vdW-DF studies of large biomolecular interaction
problems by first mapping out what interaction geometries are plausible, from knowing the
variation in the vdW attraction. By beginning the vdW-DF identification of optimal interaction 
geometries with this vdW-attraction step we are pursuing a course that is similar to that 
Nature uses in its own molecular-recognition search.

\acknowledgments

The authors thank Professor Jason Kahn, University of Maryland, for 
valued discussions on the atomic coordinates in a standard idealized  
model of structure of a ds-DNA coil.\cite{SB1,SB2} We thank Eskil Varenius 
and Magnus Sand\'en for computing a corresponding ds-B-DNA electron density 
that represents a starting point for the here-described mapping of
the vdW  attraction.  Partial support from the Swedish Research Council (VR) 
and the Chalmers Area of Advance Materials is gratefully acknowledged.
The computations were performed on resources provided by the Swedish
National Infrastructure for Computing (SNIC) at C3SE 
 and NSC. 

\appendix*

\section{Scaling of interaction at large separations}

With some robust assumptions  it is possible to provide
analytical results for the vdW interaction per length, $E_A/L$, for
two parallel (infinitely repeated) cylindrical structures when these
are far, but not asymptotically far, apart.\cite{CNTparallelconcentric,kleis_cms2005,kleis_prb2008}
The assumptions are that the matter is continuous, a slightly simplified 
plasmon-pole approximation, and that the (screened) effective susceptibility can 
be approximated as having equivalent magnitudes for response along the cylinder axis and 
tangent.  There are some differences in this analytical evaluation for hollow 
(relevant for carbon nanotubes) and filled (relevant for DNA) cylinders but in 
both cases the results can be expressed in terms of a generalized hypergeometric 
function.  For either types of interacting systems, we consider two cylinders of 
equal radius $b$ at center-to-center distance $d$.

\textit{For a pair of massive cylinders} a two-variable expression is given in the book by Mahanty and 
Ninham\cite{mahanty,comment} in terms of Appell’s hypergeometric function $F_4$
\begin{eqnarray}
\frac{E^\filled_A}{L} &=& -B \frac{b^4}{d^5}\,
F_4\left(\frac{5}{2},\frac{5}{2};2,2;\frac{b^2}{d^2},\frac{b^2}{d^2} \right) \\
&=& 
-B \frac{b^4}{d^5} \sum_{m,n=0}^{\infty}
\left( \frac{\Gamma\left(\frac{5}{2}+m+n\right)}{\Gamma\left(\frac{5}{2}\right)}\right)^2
\nonumber \\
&&
\times\frac{\left(\frac{b^2}{d^2}\right)^{m+n}}{m!\,n!\,(m+1)!\,(n+1)!}.
\end{eqnarray}
Here $B$ is a (positive) prefactor that reflects the susceptibility 
of the material in the cylinders and whose value can be set by investigating the
asymptotic form.  

We rewrite this expression with the generalized hypergeometric 
function ${}_3F_2$ of one variable
\begin{equation}
\frac{E^\filled_A}{L} = -B \frac{b^4}{d^5}\,
{}_3F_2\left(\frac{3}{2},\frac{5}{2},\frac{5}{2};2,3;4\frac{b^2}{d^2}\right).
\end{equation}
For large separations $d\gg b$ an expansion in $b/d$ yields
\begin{eqnarray}
\lefteqn{{}_3F_2\left(\frac{3}{2},\frac{5}{2},\frac{5}{2};2,3;4\frac{b^2}{d^2}\right)=}
\nonumber \\
&& 1 + \frac{25}{4} \frac{b^2}{d^2} + \frac{6125}{192} \frac{b^4}{d^4}
+ \frac{77175}{512} \frac{b^6}{d^6} + {\cal O}\left(\frac{b}{d}\right)^7
\label{eq:expanfilled}
\end{eqnarray}
which is the expansion used in the bottom panel of Figure \ref{fig:X5} 
for the pair of DNA (approximated as filled cylinders).

\textit{For a pair of infinitely thin, hollow cylinders,} where the electron density
can be described by a radial $\delta$-function, the interaction is given by a similar
generalized hypergeometric function\cite{CNTparallelconcentric,kleis_cms2005}
\begin{equation}
\frac{E^\hollow_A}{L} =
-B\frac{b^4}{d^5}\,
{}_3F_2\left(\frac{1}{2},\frac{5}{2},\frac{5}{2};1,1;4\frac{b^2}{d^2}\right)\,.
\label{eq:hollowall}
\end{equation}

The result is also valid for a slightly more general choice of 
susceptibility tensors.\cite{CNTparallelconcentric,kleis_cms2005}
For the function (\ref{eq:hollowall}) the expansion in $b/d$ yields
\begin{eqnarray}
\lefteqn{{}_3F_2\left(\frac{1}{2},\frac{5}{2},\frac{5}{2};1,1;4\frac{b^2}{d^2}\right)=}
\nonumber \\
&& 1 + \frac{25}{2} \frac{b^2}{d^2} + \frac{3675}{32} \frac{b^4}{d^4}
+\frac{55125}{64} \frac{b^6}{d^6} + {\cal O}\left(\frac{b}{d}\right)^7.
\label{eq:expanhollow}
\end{eqnarray}
This expansion is used in the bottom panel of Figure \ref{fig:X5}
for the pair of CNT (approximated as pipes, that is, hollow cylinders).



\begin{thebibliography}{99}

\bibitem{SchrodingerLife} 
E. Schr\"odinger, \textit{What is life?}
(Cambridge University Press, Cambridge, 1967).
 
\bibitem{FrancisAndCrickSet} 
J.D.~Watson and F.H.C. Crick, 
Nature \textbf{171}, 4356 (1953).

\bibitem{RutgersGenome}
H.M. Berman, W.K. Olson, D.L. Beveridge, J. Westbrook, A. Gelbin,
T. Demeny, S.-H. Hsieh, A.R. Srinivasan, and B. Schneider,
Biophys. J. \textbf{63}, 751 (1992);
http://ndbserver.rutgers.edu 

\bibitem{MolRegBook} 
M.C. Williams, L. J. Maher III, eds., 
``Biophysics of DNA-Protein Interactions" (Springer, New York, 2011).

\bibitem{MolRecognition1} 
A.L. Lehninger, D.L. Nelson, and M.M. Cox, 
Principles of Biochemistry, (Worth Publishers, Inc., New York, 1993).

\bibitem{MolRecognition2} 
For example, S.H. Gellman, Chem. Rev. \textbf{97}, 1231 (1997) 
and associated collected reviews in 
Chem. Rev. \textbf{97}; 
P.J. Edmonson \textit{et al.}, 
Int.~J.~Mod.~Sci.~\textbf{9}, 154 (2008).

\bibitem{dna1}
V.R.~Cooper, T.~Thonhauser, A. Puzder, E. Schr\"oder, B.I. Lundqvist, 
and D.C. Langreth, J. Amer. Chem. Soc. \textbf{130}, 1304 (2008).

\bibitem{dna2} 
S.~Li, V.R. Cooper, T. Thonhauser, B.I. Lundqvist, and D.C. Langreth,
J.~Phys.~Chem.~B \textbf{113}, 11166 (2009).

\bibitem{BurkePerspectives} 
K.~Burke, J.~Chem.~Phys.~\textbf{136}, 150901 (2012).

\bibitem{HardNumbers} 
H.~Rydberg, N.~Jacobson, P.~Hyldgaard, S.I.~Simak, B.I.~Lundqvist, and D.C.~Langreth,
Surf.~Sci.~\textbf{532--535}, 606 (2003). 

\bibitem{ijqc}
D.C. Langreth, M. Dion, H. Rydberg, E. Schr\"oder, P. Hyldgaard, and
B.I. Lundqvist, Intern. J. Quant. Chem. \textbf{101}, 599 (2005).

\bibitem{langrethjpcm2009}
D.C.~Langreth, B.I.~Lundqvist, S.D.~Chakarova-K\"ack, V.R.~Cooper,
M.~Dion, P.~Hyldgaard, A.~Kelkkanen, J.~Kleis, L.~Kong, S.~Li, P.G.~Moses,
E.~Murray, A.~Puzder, H.~Rydberg, E.~Schr\"oder, and T.~Thonhauser,
J.~Phys.: Cond.~Matter \textbf{21}, 084203 (2009).

\bibitem{Dion}
M.~Dion, H.~Rydberg, E.~Schr\"oder, D.C.~Langreth, and B.I.~Lundqvist,
Phys. Rev. Lett. \textbf{92}, 246401 (2004); \textbf{95}, 109902(E) (2005).

\bibitem{Thonhauser}
T. Thonhauser, V.R. Cooper, S. Li, A. Puzder, P. Hyldgaard, and D.C. Langreth,
Phys.~Rev.~B \textbf{76}, 125112 (2007).

\bibitem{vdWDF2}
K.~Lee, E.D.~Murray, L.~Kong, B.I.~Lundqvist, and D.C.~Langreth,
Phys.~Rev.~B \textbf{82}, 081101(R) (2010).

\bibitem{MolCrys}
K. Berland and P. Hyldgaard,
J. Chem. Phys. \textbf{132}, 134705 (2010).

\bibitem{MolCrys2}
K. Berland, {\O}. Borck, and P. Hyldgaard,
Comp. Phys. Commun. \textbf{182}, 1800 (2011).

\bibitem{ourharris}
K.~Berland, E. Londero, E. Schr\"oder, and P. Hyldgaard,
\textit{A Harris-type van der Waals density functional scheme}, 
arXiv:1303.3762  [cond-mat.mtrl-sci].

\bibitem{EarlyLength} 
Y.U. Barash, 
Fiz.~Tverd.~Tela.~\textbf{30}, 1578 (1988);
B. E. Sernelius and P. Bj\"ork, 
Phys. Rev. B \textbf{57}, 6592 (1998).

\bibitem{DobsonReview} 
J.~F.~Dobson and T.~Gould, 
J.~Phys.: Condens.~Matter \textbf{24}, 073201 (2012);
S.~Leb\'egue, J.~Harl, T.~Gould, J.~\'Angy\'an, G.~Kresse, and J.F.~Dobson, 
Phys.~Rev.~Lett.~\textbf{105}, 196401 (2010).

\bibitem{JPlength} 
J.P.~Perdew, J.~Tao, P.~Hao, A.~Ruzsinszky, G.I.~Csonka, and J.M.~Pitarke, 
J.Phys.: Condens.~Matter \textbf{24}, 424207 (2012);
A.~Ruzsinszky, J.P.~Perdew, J.~Tao, G.I.~Csonka, and J.~M.~Pitarke, 
Phys.~Rev.~Lett.~\textbf{109}, 233203 (2012).

\bibitem{soler}
G.~Rom\'an-P\'erez and J.M.~Soler,
Phys.~Rev.~Lett. \textbf{103}, 096102 (2009).

\bibitem{KGraphite}
E.~Ziambaras, J.~Kleis, E.~Schr{\"o}der, and P.~Hyldgaard,
Phys.~Rev.~B~\textbf{76}, 155425 (2007).

\bibitem{Gulans} 
A.~Gulans, M.J.~Puska, and R.M.~Nieminen,
Phys.~Rev.~B \textbf{79}, 201105(R) (2009).

\bibitem{noloco}
D.~Nabok, P. Puschnig, C. Ambrosch-Draxl,
Comp. Phys. Commun. \textbf{182}, 1657 (2011).

\bibitem{Junolo}
Open-source tool \textsc{JuNoLo} for real-space and fast-fourier transform
non-selfconsistent vdW-DF total-energy evaluation;
P.~Lazi{\'c}, N.~Atodiresei, M.~Alaei, V.~Caciuc, S.~Bl{\"u}gel,
and R.~Brako, Comp.~Phys.~Comm. \textbf{181}, 371 (2010).

\bibitem{kleis_prb2008}
J. Kleis, E. Schr\"oder, and P. Hyldgaard,
Phys. Rev. B \textbf{77}, 205422 (2008).

\bibitem{DFTDset} 
X.~Wu, M.C.~Vargas, S.~Nayak, V.~Lotrich, and G.~Scoles,
J. Chem. Phys. 115, 8748 (2001); 
S.~Grimme, 
J.~Comp.~Chem.~\textbf{25}, 1463 (2004).

\bibitem{Grimme} 
S.~Grimme, J.~Comp.~Chem. \textbf{27}, 1787 (2006).

\bibitem{concentric}
E. Schr\"oder and P. Hyldgaard,
Surf. Sci. \textbf{532}, 880 (2003).
 
\bibitem{CNTparallelconcentric}
E. Schr\"oder and P. Hyldgaard,
Mater. Sci. Engin. C \textbf{23}, 721 (2003).

\bibitem{kleis_cms2005}
J. Kleis, P. Hyldgaard, and E. Schr\"oder,
Comp. Mater. Sci. \textbf{33}, 192 (2005).

\bibitem{harris}
J. Harris, Phys. Rev. B \textbf{31}, 1770 (1985).

\bibitem{foulkes}
W.M.C. Foulkes and R. Haydock,
Phys. Rev. B \textbf{39}, 12520 (1989).

\bibitem{nikulin}
V.K. Nikulin, Zh. Tekhn. Fiz. \textbf{XLI}, 41 (1971)
[Sov. Phys. - Techn. Phys. \textbf{16}, 28 (1971)].

\bibitem{gordonkim}
R.G. Gordon and Y.S. Kim, J. Chem. Phys. \textbf{56}, 3122 (1972).

\bibitem{xcrysden}
A. Kokalj, Comp. Mater. Sci. \textbf{28}, 155 (2003).
Code available from http://www.xcrysden.org/


\bibitem{SB1}
S. Arnott and D.W.L. Hukins,
Biochem. Biophys. Research Commun. \textbf{47}, 1504 (1972).

\bibitem{SB2}
S. Arnott, D.W.L. Hukins, and S.D. Dover,
Biochem. Biophys. Research Commun. \textbf{48}, 1392 (1972).

\bibitem{Jason}
The idealized atomic coordinates for the here-investigated
base-pair sequence in our infinitely repeated ds-DNA model system
were put together by Prof. Jason Kahn, University of Maryland. 
The coordinates are available as  
part of his instructive rasmol tutorial, available
at http://www.biochem.umd.edu/biochem/kahn/teach\_res

\bibitem{gpaw}
Open-source, grid-based PAW-method DFT code \textsc{gpaw},
http://wiki.fysik.dtu.dk/gpaw/

\bibitem{newvdWanalysis}
K.~Berland and P.~Hyldgaard,
\textit{An analysis of van der Waals density functional components:
Binding and corrugation of
benzene and C60 on boron nitride and graphene},
arXiv:1303.0389 [cond-mat.mes-hall].

\bibitem{adenine}
K. Berland, S.D. Chakarova-K\"ack, V.R. Cooper, D.C. Langreth, and E. Schr\"oder,
J. Phys.: Cond. Matt. \textbf{23}, 135001 (2011).

\bibitem{kleis_jcp2005}
J. Kleis and E. Schr\"oder,
J. Chem. Phys. \textbf{122}, 164902 (2005).

\bibitem{kleis_prb2007}
J. Kleis, B.I. Lundqvist, D.C. Langreth, and E. Schr\"oder,
Phys. Rev. B  \textbf{76}, 100201(R) (2007).

\bibitem{mahanty}
J. Mahanty and B.W. Ninham,
\textit{Dispersion Forces\/} 
(Academic Press, London, 1976).

\bibitem{comment}
Please note that there is a typographical error
in Ref.\ \onlinecite{mahanty} in what corresponds to
the last expression below.

\end{thebibliography}
\end{document}